\documentclass[a4paper,11pt]{article}
\pdfoutput=1

\usepackage{jheppub}
\usepackage{subfigure}
\usepackage[T1]{fontenc}
\usepackage[english]{babel}
\usepackage[applemac]{inputenc}
\usepackage{mathtools}
\usepackage{booktabs}
\usepackage{slashed}
\usepackage{float}
\usepackage{tabularx}
\newcommand{\be}{\begin{equation}}
\newcommand{\ee}{\end{equation}}
\newcommand{\nn}{\nonumber}

\title{Muonic vs electronic dark forces: a complete EFT treatment for atomic spectroscopy}
\author[a]{Claudia Frugiuele,}
\author[b]{and Clara Peset}

\affiliation[a]{INFN, Sezione di Milano,\\ Via Celoria 16, \\I-20133 Milano, Italy}
\affiliation[b]{Technische Universit\"at M\"unchen,\\
Physik Department,\\
James-Franck-Stra\ss e 1,\\
D-85748 Garching, Germany}

\emailAdd{claudia.frugiuele@cern.ch}\emailAdd{clara.peset@tum.de}

\abstract{Precision atomic spectroscopy provides a solid model independent bound on the existence of new dark forces among the atomic constituents.
We focus on the keV-GeV region investigating the sensitivity to such dark sectors of the recent measurements on muonic atoms at PSI. To this end we develop for the first time, the effective field theory that describes the leading effect of a new (pseudo-)vector or a (pseudo-)scalar particle of any mass at atomic energies.
We identify in  the Lamb Shift  measurement in muonic deuterium ($\mu$D) and the $2s$ Hyperfine Splitting (HFS) in muonic hydrogen ($\mu$H) the most promising measurements to probe respectively spin-independent  and spin-dependent new forces. Furthermore, we evaluate the expression of the vector force HFS finding that a future measurement of the $2s$ HFS in regular hydrogen could provide the strongest atomic bound for such a force for masses above 100 MeV.
\par
\begin{flushright}
TUM-HEP-1354/21
\end{flushright}
}

\begin{document}

\maketitle
\baselineskip=18pt

\section{Introduction}

There have been substantial advances in controlling matter and light in the last 20 years, creating new opportunities in atomic, molecular and optical physics such as  novel probes of physics beyond the Standard Model (BSM) \cite{budkerreview,Derevianko:2013oaa,Jaeckel:2010xx,Karshenboim:2011dx,Karshenboim:2010cg,helium,exotic,Antypas:2020rtg,Antypas:2019qji,Jones:2019qny,kings,kings1,atomic1}.  This has come at the ideal time for particle physics. 
Indeed, traditional approaches to BSM physics have been deeply challenged by the results of new physics searches carried out at the Large Hadron Collider (LHC) \cite{Giudice:2017pzm}, and by the absence of a positive signal at dark matter (DM) direct detection experiments.
As a reaction, new ideas have blossomed showing  how BSM compelling inquiries such as the existence of DM \cite{uscosmic} or the hierarchy problem \cite{relaxion} could be connected to the existence of  low mass dark sectors,  very weakly coupled to the visible sector. Since high energy colliders have limited sensitivity to such dark particles, the main players in this quest are experiments  at the intensity or precision frontier \cite{PBC}.
\par
\par
New forces between the dark and the visible sectors might modify atomic energy levels and thus could be detected via precision spectroscopy. For simple atoms such as atoms with few electrons or exotic hydrogen-like atoms, one can rely on the comparable precision between the theoretical calculations and the experimental measurements such that the effect of dark forces can be bounded by comparing the experimental and theoretical errors  \cite{Jaeckel:2010xx,Karshenboim:2011dx,Karshenboim:2010cg,helium,exotic}. 
For atoms with many electrons, a different strategy \cite{kings,kings1,atomic1} should be followed due to the limited precision of the theoretical calculations.
\par
In this work, we will examine the sensitivity to dark forces of muonic atoms, that is hydrogen-like atomic bound systems formed by a negative muon and a nucleus. These atoms are are smaller than standard atoms by a factor of $\left(m_e/m_{\mu}\right)^3 \sim 8\cdot 10^6$, which makes them potentially more sensitive to intermediate scale new forces, i.e. mediated by particles with masses in the MeV-GeV range. 
 Such forces are typically mediated by particles with masses close to the reduced mass of the system $\sim m_\mu$ which typically produce a contact interaction rather than a potential-like exchange at atomic-like energies. We deal with this issue by developing, for the first time, an effective field theory (EFT) that accounts for the leading effect of mediators in any mass range. We also compute the one-loop matching coefficients that some of the leading effects which we study here require. 
The paper is organized as follows.
In Section~\ref{sec:LS}, we present the contribution given by new force carriers to the Lamb shift (LS) of a generic hydrogen-like atom. In Section~\ref{sec:HF}, we consider the HFS not only for spin-dependent, but also for spin-independent forces mediated by a new vector boson. In Section~\ref{sec:1s2s}, we provide expressions for $1s-2s$ splittings for atoms with components of equal and different masses.
In Section~\ref{sec:muonicspectroscopy}, we explore the sensitivity of the PSI measurements on muonic atoms to spin-independent and spin-dependent dark forces. We also update the bound from the Lamb shift measurement in regular hydrogen considering the recent experimental result \cite{Bezginov:2019mdi} and the new determination of the proton charge radius.
\section{Testing dark forces  via Lamb shift measurements}\label{sec:LS}

We consider new physics forces mediated by an axial-vector ($A$), vector ($V$), scalar ($S$) or pseudo-scalar ($S$) particle of mass $m_\phi$ that couples to the Standard Model fermions as 
\begin{align}\label{eq:NPLags}
\mathcal{L}_V=g_V\bar{\psi}\slashed{V}\psi,\quad\mathcal{L}_A=g_A\bar{\psi}\slashed{A}\gamma^5\psi,\quad\mathcal{L}_S=g_S\bar{\psi}S\psi,\quad\mathcal{L}_P=ig_P\bar{\psi}P\gamma^5\psi.
\end{align}
We take an EFT perspective to organize the contributions of the different force carriers to atomic energy levels. The advantage of this approach is threefold: first, we can disentangle the leading contribution in an efficient and minimal way, thus simplifying computations. Second, we work in a framework in which we have control over the origin and nature of divergences and their associated counterterms. On top of this, we can keep track of the size of the uncomputed terms, thus allowing for a robust error determination.

The potential nonrelativistic EFT that describes the effect of new force carriers in energy shifts is reached by organizing the different scales in the problem. Bound states are typically nonrelativistic systems in which there are three clearly differentiated energy regimes:
\begin{itemize}
\item Hard sale: $m_r=\tfrac{m_1 m_2}{m_1+m_2}\sim m_1\sim m_2$,
\item Soft scale: $a_0^{-1}=m_rZ\alpha\sim |{\bf p}|\sim 1/r$,
\item Ultrasoft scale: $m_r\alpha^2\sim E$,
\end{itemize}
where $\alpha$ is the hyperfine structure constant, $m_{1,2}$ are the masses of the bound state components, $m_r$ its reduced mass, $E$ its typical binding energy and ${\bf p}=m_r v$ ($r$) the typical momentum (size). In a bound system, the hard and soft scales can be integrated out. The resulting EFT naturally provides a Schr\"odinger-like formulation of the bound-state problem while keeping the quantum field theory nature of the interaction with ultrasoft degrees of freedom, and introducing the information of the high-energy modes (also of a quantum field theory nature) in the Wilson coefficients of the theory \cite{Pineda:1997bj,Pineda:1998kn,Peset:2015zga}.
Given the large amount of precision tests of QED, the contribution of the new force cannot be large compared to the contribution of photons. Therefore, we consider it to be a perturbation to the Coulomb potential, rather than solving the Schr\"odinger equation exactly. The latter is equivalent to a resummation of higher order effects which will overall amount to a small perturbation of the leading effect, while breaking the power counting of the EFT.

The way in which the new physics effects enter the EFT will depend on the size of $m_\phi$. We will distinguish two different EFTs: one where $m_\phi$ is light, i.e. of the order of the soft scale or smaller, and another where it is heavy, i.e. it scales like the bound state masses (hard scale). We will see that the matching between these two theories is smooth, as it should be. The full EFT for the force carriers in Eq.~\eqref{eq:NPLags} is presented in Appendix~\ref{app:EFTforNP}.

 The Lamb shift in atoms with components of different mass is given by the energy difference of the $2p$ and $2s$ spin-averaged states $E_\text{LS}=E(2p_{1/2})-E(2s_{1/2})$. This means that only spin-independent interactions like those of vectors and scalars will contribute.\footnote{This is not the case for particle-antiparticle states such as positronium, where measurements of non-spin averaged transitions are carried out, and thus get contributions from spin-dependent forces. This was studied in detail in \cite{Frugiuele:2019drl}.}

The leading order potentials from a new vector \cite{Holdom:1985ag,atomic1} or scalar exchange \cite{relaxion}, contributing to the LS can be expressed in a compact way as ($X=V,S$)
\begin{align}
V_\text{LS}(r)=\begin{dcases}
      c_X \frac{g_X^{(1)}g_X^{(2)}}{4\pi r}e^{-m_\phi r} & \quad \text{ for }m_\phi\lesssim  a_0^{-1},  \\
     \frac{d_S^{(X)}}{m_1 m_2}\delta^{(3)}(r) & \quad \text{ for }m_\phi\sim m_r ,
  \end{dcases} 
\end{align}
where $c_V=1$, and $c_S=-1$. 
The energy shift produced by this potential can be obtained with the expectation values compiled in Appendix~\ref{app:EVs}, which gives
\begin{align}\label{eq:ELS}
E_\text{LS}=\begin{dcases}
      -\frac{c_X g_X^{(1)}g_X^{(2)}a_0 m_\phi^2}{8\pi (1+a_0 m_\phi)^4} & \quad \text{ for }m_\phi\lesssim  a_0^{-1} , \\
     -\frac{c_X g_X^{(1)}g_X^{(2)}}{8\pi a_0^3 m_\phi^2} & \quad \text{ for }m_\phi\sim m_r ,
  \end{dcases} 
\end{align}
where $a_0=1/(m_r Z \alpha)$ is the atomic Bohr radius. In \eqref{eq:ELS}, the leading correction to both regimes of $m_\phi$ comes from a tree level exchange, and so the upper relation is equivalent to resuming all the series expansion in $1/m_\phi$ in the heavy mass theory. 
Note that for $Z>1$, $g_X^{(N)}= Z g_X^{(p)}$.
\par
We observe that a scalar particle gives an opposite sign contribution with respect to a vector dark force for a fermion-fermion potential.
Hence, we could in principle envision a scenario where the LS contributions are canceled among different particles.
Note that this is different with respect the $g-2$ case where cancellations are possible without the necessity to add new matter \cite{Giudice:2012ms}.
Furthermore, as we will see in the next section, combining HFS and LS measurements provide a test for dark forces free from possible cancellations. The reason is that a vector force gives a non-zero contribution to the HFS, while scalars do not.

\section{Testing  dark forces  via Hyperfine splitting measurements}\label{sec:HF}

We consider the hyperfine splitting of the $1s$ states $E_\text{HF}=E(1s^{f=1})-E(1s^{f=0})$. In this case, only the light axial-vector contributes to this splitting at leading order in the non-relativistic expansion (${\bf p}/m_r$)
\begin{align}\label{eq:VHFA}
V_\text{HF,A}(r)=\begin{dcases}
     -\frac{g_A^{(1)}g_A^{(2)}}{\pi r}{\bf S_1}\cdot{\bf S_2} & \quad \text{ for }m_\phi\lesssim  a_0^{-1},  \\
   -\frac{4d_v^{(A)}}{m_1m_2}\delta^{(3)}(r){\bf S_1}\cdot{\bf S_2} & \quad \text{ for }m_\phi\sim m_r ,
  \end{dcases} 
\end{align}
where ${\bf S_i}$ refers to the spin of particle-$i$ and $d_v^{(A)}$ can be found in \eqref{dvAm}.

The energy shift produced by this potential is again computed with the results in Appendix~\ref{app:EVs} and gives
\begin{align}
E_\text{HF,A}=\begin{dcases}
-\frac{4g_A^{(1)}g_A^{(2)}}{\pi a_0(2+a_0m_\phi)^2}& \quad \text{ for }m_\phi\lesssim m_r\alpha\sim a_0^{-1},  \\
-\frac{4g_A^{(1)}g_A^{(2)}}{\pi a_0^3 m_\phi^2} & \quad \text{ for }m_\phi\sim m_r ,
\end{dcases}
\end{align}
where again, as was the case for the Lamb shift, the heavy-$m_\phi$ limit of the expression for light mass naturally reproduces the heavy mass one, since both come from tree level exchange.

There is no contribution form a scalar to the hyperfine splitting up to $\mathcal{O}({\bf p}^2/m_r^2)$, while the vector and pseudo-scalar contributions can be compactly expressed by the potential ($X=V,P$)
\begin{align}\label{eq:VHF}
V_\text{HF}(r)=\begin{dcases}
    h_X\frac{g_X^{(1)}g_X^{(2)}}{6\pi m_1m_2}\left(\frac{m_\phi^2e^{-m_\phi r}}{r}-4\pi\delta^{(3)}(r)\right){\bf S_1}\cdot{\bf S_2} & \quad \text{ for }m_\phi\lesssim  a_0^{-1} , \\
    -\frac{4d_v^{(X)}}{m_1m_2}\delta^{(3)}(r){\bf S_1}\cdot{\bf S_2} & \quad \text{ for }m_\phi\sim m_r ,
  \end{dcases}
\end{align}
with $h_V=1, h_S=0, h_P=-\frac{1}{2}$ and where now $d_v^{(V)}$ can be found in \eqref{dvVm} and $d_v^{(P)}$ in\eqref{dvPm}.

\begin{figure}
\begin{center}
\begin{tabular}{lccc}
&QED+NP && NREFT\\\\
\raisebox{3.5\height}{Light mass NP}&\includegraphics[width=.2\textwidth]{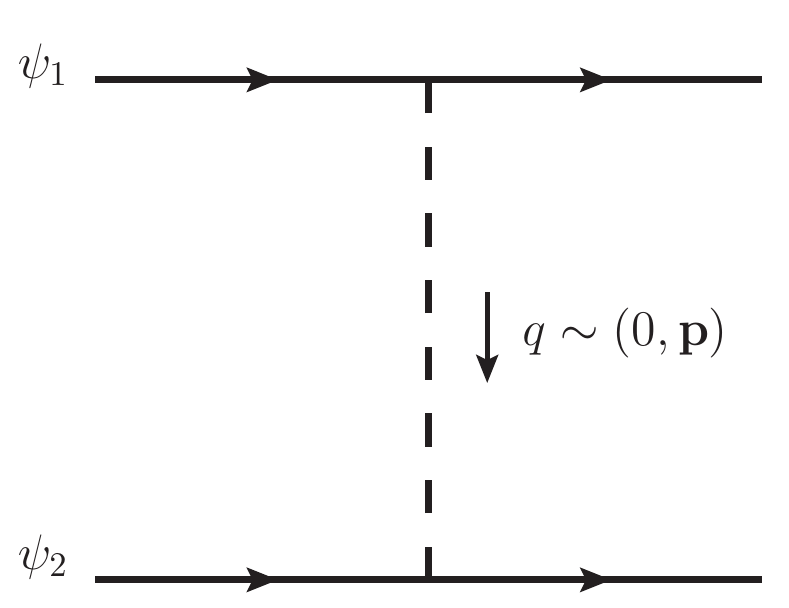}&\raisebox{6\height}{$\Rightarrow$} &\includegraphics[width=.2\textwidth]{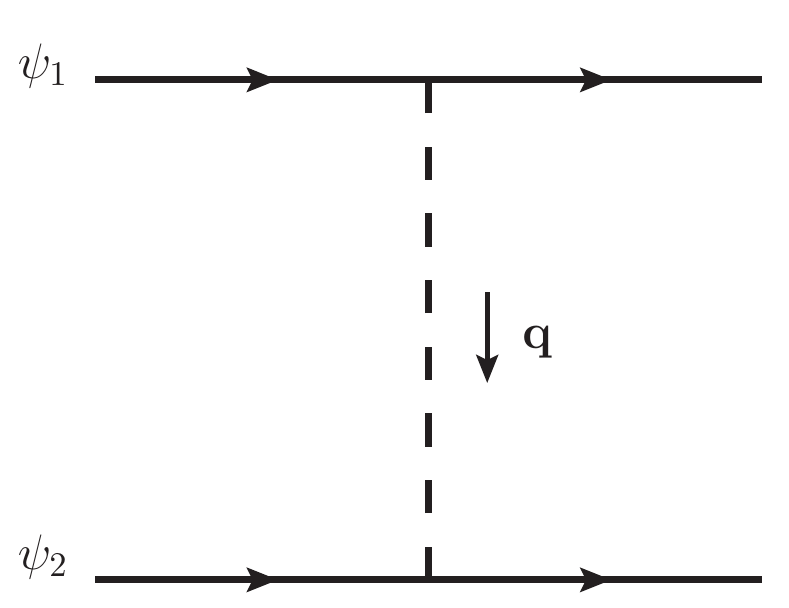}\\\\
\raisebox{3.5\height}{Heavy mass NP}&\includegraphics[width=.4\textwidth]{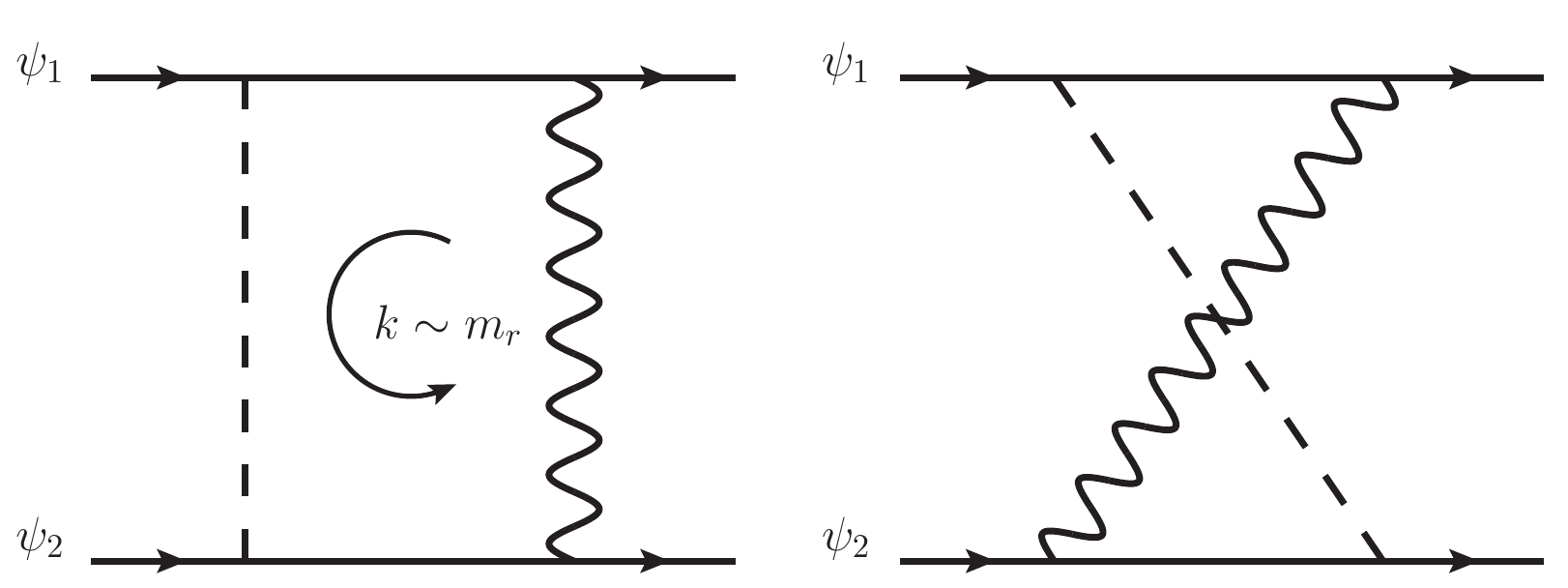}&\raisebox{6\height}{$\Rightarrow$}&\includegraphics[width=.2\textwidth]{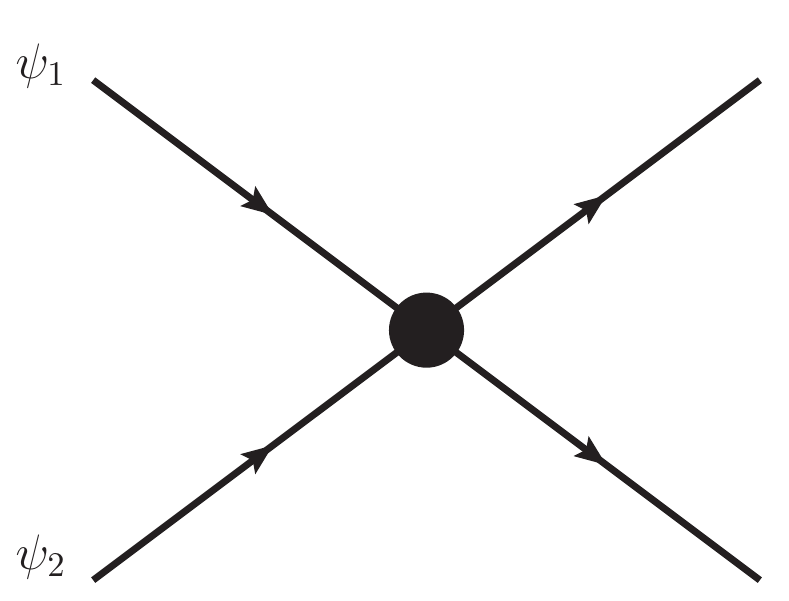}
\end{tabular}
\end{center}
\caption{Graphic representation of the matching between the full theory and the NREFT for the HF pseudo-scalar potential in the light and heavy mass theories. Continuous lines represent the fermionic components of the bound state, while dashed and wiggly lines represent the pseudoscalar particle and the photon respectively.}\label{FigNREFT}
\end{figure}

The vector and pseudo-scalar the energy shift is then
\begin{align}
E_\text{HF}=\begin{dcases}
-h_X\frac{8g_X^{(1)}g_X^{(2)}}{3\pi a_0^3 m_1m_2}\frac{1+a_0m_\phi}{(2+a_0m_\phi)^2}& \quad \text{ for }m_\phi\lesssim m_r\alpha\sim a_0^{-1},  \\
-\frac{4d_v^{(X)}}{\pi a_0^3 m_1m_2} & \quad \text{ for }m_\phi\sim m_r.
\end{dcases}
\end{align}
Now, the energy shift in the light mass EFT comes from a tree-level exchange, while the contribution from the heavy mass EFT arises from the matching onto the one-loop diagrams in the full theory, as graphically described Fig.~\ref{FigNREFT}. This produces a non-trivial matching between the heavy and light mass theories
\begin{align}
\lim_{m_\phi\to\infty}E_\text{HF}\bigg|_\text{light}=\lim_{m_\phi\to 0}E_\text{HF}\bigg|_\text{heavy}=-h_X\frac{8g_X^{(1)}g_X^{(2)}}{3\pi a_0^4m_1m_2m_\phi}.
\end{align}
thus providing a theory for all the possible range of $m_\phi$.

The interpretation of this finding is that the leading contribution for a heavy vector/pseudoscalar is not a tree level contribution but rather a contribution at one-loop. This becomes clear if we consider the tree-level potential in momentum space for a heavy pseudoscalar mediator
\begin{align}
\tilde V({\bf q})=\frac{g_P^{(1)}g_P^{(2}}{4m_1m_2 m_\phi^2}(\boldsymbol{\sigma}_1 {\bf q}
)(\boldsymbol{\sigma}_2 {\bf q})
\end{align}
which produces a position space potential $\sim 1/r^5$ with a divergent expectation value. This hints the presence of a less suppressed contribution with a finite expectation value that can then provide the counterterm to regularize the tree-level divergent contribution. This is a clear example of the power of using EFTs to organize computations.
The contribution of the heavy mass vector/pseudoscalar mediators to the HFS is a new result. It extends the computation for the vector case in \cite{Karshenboim:2014tka}, which considers only the extremely heavy case $m_\phi\gg m_1,m_2$.

The results for the ground-state can be easily extended to the $2s$ HF splitting, $E_\text{HF2}=E(2s^{f=1})-E(2s^{f=0})$.
The energy shift produced by the HF axial potential in \eqref{eq:VHFA} is
\begin{align}
E_\text{HF2,A}=\begin{dcases}
-\frac{g_A^{(1)}g_A^{(2)}}{4\pi }\frac{1+2a_0^2m_\phi^2}{a_0(2+a_0m_\phi)^4}& \quad \text{ for }m_\phi\lesssim m_r\alpha\sim a_0^{-1},  \\
-\frac{g_A^{(1)}g_A^{(2)}}{2\pi a_0^3 m_\phi^2} & \quad \text{ for }m_\phi\sim m_r,
\end{dcases}
\end{align}
and for the vector and pseudoscalar in \eqref{eq:VHF}
\begin{align}
E_\text{HF2}=\begin{dcases}
-h_X\frac{g_X^{(1)}g_X^{(2)}}{24\pi a_0^3 m_1m_2}\frac{a_0 m_\phi (a_0 m_\phi (8 a_0 m_\phi+11)+8)+2}{(a_0 m_\phi+1)^4}& \quad \text{ for }m_\phi\lesssim m_r\alpha\sim a_0^{-1},  \\
-\frac{d_v^{(X)}}{2\pi a_0^3 m_1m_2} & \quad \text{ for }m_\phi\sim m_r .
\end{dcases}
\end{align}

\section{Testing dark forces via $1s-2s$ splitting measurements}\label{sec:1s2s}
The $1s-2s$ splitting is typically given by the energy difference $2^{2f+1}s-1^{2f+1}s$  with $f=0,1$, in systems of equal mass components such as positronium or true muonium, or $2s_{1/2}^{f}-1s_{1/2}^{f}$  with $f=0,1$, in atoms with different mass components, such as hydrogen-like atoms or muonium. Since this energy difference does not involve spin-averaged states, it gets contributions both from the LS and the HF potentials described in the previous sections. 

For the vector, scalar and axial particles, the potential is the leading order Yukawa-like one and we obtain,
\begin{align}
E_\text{1S2S}=\begin{dcases}
f_X^{(s)}\frac{ g_X^{(1)} g_X^{(2)} }{16 \pi  a_0}\left(\frac{2 a_0^2 m_\phi^2+1}{(a_0 m_\phi+1)^4}-\frac{16}{(a_0 m_\phi+2)^2}\right)& \quad \text{ for }m_\phi\lesssim m_r\alpha\sim a_0^{-1},  \\
-f_X^{(s)}\frac{ 7 g_X^{(1)} g_X^{(2)}}{8 \pi  a_0^3 m_\phi^2} & \quad \text{ for }m_\phi\sim m_r, 
\end{dcases}
\end{align}
with $f_V^{(0)}=f_V^{(1)}=1$, $f_S^{(0)}=f_S^{(1)}=-1$, $f_A^{(0)}=3$, $f_A^{(1)}=-1$. 
The case of the pseudoscalar is again more interesting as the potential starts at $\mathcal{O}(1/m_r^2)$
\begin{align}
E_\text{1S2S,P}=\begin{dcases}
\frac{(3-4s)g_P^{(1)} g_P^{(2)}  }{192 \pi  a_0^3 m_1m_2}\left(14-\frac{16 a_0^2 m_\phi^2}{(a_0 m_\phi+2)^2}+\frac{a_0^2 m_\phi^2 \left(2 a_0^2 m_\phi^2+1\right)}{(a_0 m_\phi+1)^4}\right)
&  \text{for }m_\phi\lesssim m_r\alpha, \\
\frac{-7(3-4s)d_v^{(P)}}{8\pi a_0^3 m_1m_2} &  \text{for }m_\phi\sim m_r. 
\end{dcases}
\end{align}
There are currently no competitive measurements of the $1s-2s$ splitting in muonic atoms, although they are expected to be a part of the new era of muonic experiments.

\section{Probing dark forces via laser muonic spectroscopy }\label{sec:muonicspectroscopy}
Muonic atoms are exotic hydrogen-like atoms where an electron is replaced by a muon.
The spectroscopy of muonic atoms offers an interesting opportunity to extract properties of the nucleus with high accuracy. The reason is that the muon is lighter than the electron, and hence its  wave function has a larger overlap with the nucleus. 
X-ray spectroscopy of muonic atoms has for a long time been used to determine absolute
charge radii for nuclei above carbon \cite{vacchi}.
For lighter atoms instead Lamb Shift measurements, due to their weak dependence on the Rydberg constant, provide the possibility of measuring  the nucleus radius with better precision with respect to other experimental limits \cite{Peset:2021iul}, that is 
 elastic electron-nucleus scattering measurements and  high-precision laser spectroscopy in standard atoms.
This is the strategy followed by the CREMA experiment at the PSI \cite{Pohl:2010zza,Antognini:1900ns}. The first measurement carried out there was on muonic hydrogen in 2010, and led to a new  measurement of the proton charge radius one order of magnitude more precise than the previous ones.
 Nevertheless, this new, more precise method yielded a measurement  $4\%$ smaller than the world value.
 The large discrepancy led the scientific community to revise the previous determinations (from hydrogen spectroscopy and electron-proton scattering) and, in special, their error determination (see e.g. \cite{Peset:2021iul} for a recent review).
 This has been followed by newer, more precise measurements of hydrogen energy levels, which tend to confirm the smaller proton radius.\footnote{Note however that, at present, one of the new hydrogen measurements at Paris \cite{Fleurbaey:2018fih} still gives a larger value for the proton radius. Given that an independent measurement of the same energy shift at MPQ \cite{Grinin1061}, and other independent determinations \cite{Bezginov:2019mdi, Beyer:2017gug} agree with the smaller proton radius value, we consider this to be a remaining tension for some experiments, which should, nevertheless, be clarified.} Further measurements, also at PSI, of the Lamb shift in muonic deuterium \cite{Pohl1:2016xoo} and, more recently, in muonic helium-4 ions ($\mu^4$He) \cite{Krauth:2021foz} have followed. Therefore, the present experimental situation of muonic atoms is the perfect testing ground for new muon-related force carriers.
 \par
Dark forces might modify atomic energy levels and hence  for a given transition $a \to b$  
the agreement between theory and experiment would imply that:
\be
|\Delta E_{a \rightarrow b}^{\rm BSM}| < | \Delta E_{a\rightarrow  b}^{\rm exp} -\Delta E_{a \rightarrow b}^{\rm theo} |   \lesssim2 \sigma_{\rm max},
\ee
where  $\Delta E_{a\rightarrow  b}^{\rm exp} $ is the experimental measurement and  $\Delta E_{a \rightarrow b}^{\rm theo}$ the theoretical prediction, $ \sigma_{\rm max}$ represents instead the biggest source error, that is:
\be
\sigma_{\rm max}  \equiv \rm Max( \sigma_{\rm exp}, \sigma_{\rm theo}).
\ee
Let us now present the values of $\sigma_{max}$ which we will use in the next subsections to constrain dark forces.
 

  
\bigskip
\textbf{Lamb Shift measurements}

\bigskip
 \begin{table}[!htbp]
\centering
\begin{tabular}{ccc|c|c}
\toprule
System &  \multicolumn{2}{c}{Lamb shift} & \multicolumn{2}{c}{}\\
\midrule
{}   & Exp. (MHz) & Theo. (MHz) &$\sigma_{max} $  (MHz)& $\sigma_\text{proj} $   (MHz) \\
H   &909.8717(32)\cite{Bezginov:2019mdi}   &  909.8742(3) &3.2 $10^{-3}$&0.3 $10^{-3}$ \\
\bottomrule 
{}   & Exp. (meV) & Theo.  (meV)&$\sigma_{max} $  (meV)& $\sigma_\text{proj} $ (meV) \\
$\mu$H   &202.3706(23)\cite{Pohl:2010zza,Antognini:1900ns}   & 202.397(33)  &33 $10^{-3}$ & 13 $10^{-3}$    \\
$\mu$D   &202.8785(34)\cite{Pohl1:2016xoo}   &   202.869{(22)} & 22 $10^{-3}$ & 7 $10^{-3}$\\
$\mu^4$He   &   1378.521(48)\cite{Krauth:2021foz} & 1377.54(1.46)  & 1.46 & 0.46\\
\bottomrule 
\end{tabular}
\caption{Table for the theoretical and experimental predictions for the Lamb shift in different muonic systems as well as in hydrogen. See the text for references to the theoretical computations and the origin of the projected uncertainty.}\label{table:LS}
\end{table} 
Lamb Shift measurements have been performed on $\mu$H, $\mu$D and $\mu^4$He at PSI\footnote{We do not include in our analysis measurements for $\mu^3$He as they are so far preliminary \cite{TalkPohl}.} providing the most accurate value of the nucleus charge radius for these atoms. We report in Table~\ref{table:LS} both the experimental and theoretical values for these transitions including also the regular hydrogen (H) latest Lamb Shift measurement.
We consider also the projection for realistic improvements on the uncertainty of each system.
In the following we provide a brief explanation on how the theoretical predictions are obtained and how the future projections are reached. 
\begin{itemize}
\item \textbf{Hydrogen H} The theoretical prediction is taken from \cite{PhysRevA.93.022513} using the value of the proton radius from muonic hydrogen measurements \cite{Pohl:2010zza,Antognini:1900ns}. We compare it to the latest experimental result measured in \cite{Bezginov:2019mdi}. In this case, the theoretical prediction is more accurate than the experimental measurement, and so as a future projection we consider that combined experimental measurements can reach a factor 10 improvement and reach the theoretical precision. 
\item    \textbf{Muonic Hydrogen $\mu$H} For the theoretical prediction  we use the computation in \cite{Peset:2015zga} with the proton radius obtained form the latest hydrogen spectroscopic measurements in \cite{Bezginov:2019mdi,Beyer:2017gug,Grinin1061} combined (we do not consider the measurement in \cite{Fleurbaey:2018fih} which predicts a much larger value for the proton radius). We consider that this uncertainty could be improved by a factor $\sim$1/3 when the proton radius uncertainty reaches the precision of the direct determinations from muonic hydrogen. 
\item  \textbf{Muonic Deuterium $\mu$D} 
For the theoretical prediction we use the QED computations compiled in \cite{Pohl1:2016xoo} from \cite{Krauth:2015nja} together with the deuteron radius obtained from hydrogen/deuterium $1s-2s$ isotope shift measurement \cite{Parthey:2010aya,Parthey:2011lfa} corrected by the three photon exchange \cite{PhysRevA.97.062511} with the proton radius from the muonic hydrogen Lamb shift measurement \cite{Pohl:2010zza,Antognini:1900ns}, and the polarizability correction from chiral EFT computed in \cite{Hernandez:2017mof}. Here, it is the theoretical prediction that limits the accuracy. The theoretical error is fully dominated by the polarizability contribution. Recent studies that combine chiral EFT with a dispersive analysis propose an improvement in the uncertainty of almost a 50\% \cite{Acharya:2020bxf} with still room for improvement once future measurements of the deuteron electrodisintegration at the MAMI A1 and MESA experiments are carried out. We therefore consider a projected refinement of the theoretical uncertainty by a factor $\sim$1/3.
\item  \textbf{Muonic Helium $\mu^4$He} 
We consider the latest experimental measurement in \cite{Krauth:2021foz}  and the theoretical analysis in \cite{Diepold:2016cxv}. The bulk of the theoretical uncertainty is given by the poor measurement of the $^4$He radius from electron scattering experiments \cite{Sick:2008zza}. We consider a reliable improvement of this measurement that would come e.g. from the already ongoing $\mu^3$He experiment at PSI in combination with new measurements of the isotope shift \cite{Krauth_2019,PhysRevA.79.052505}. Assuming the preliminary result for $^3$He charge radius \cite{TalkPohl}, we would obtain the $^4$He charge radius with an accuracy of $\sim$0.001 fm, thus lowering the theoretical uncertainty of the Lamb shift to 0.46 meV. Actually, also the $2p$ fine structure of the $\mu^4$He has been measured in \cite{Krauth:2021foz}, but the bounds provided by this transition will be very similar in size to the ones from the Lamb shift presented here.
\end{itemize}

\bigskip
\textbf{Hyperfine Splitting measurements }

\bigskip
Table~\ref{table:HF} presents both the theoretical and experimental predictions of the hyperfine splitting  in different muonic systems as well as in hydrogen. We consider also the projection for realistic improvements on the uncertainty of each system.
 All the hyperfine experimental measurements considered here have the common characteristic that the experimental uncertainty is larger than the theoretical one, thus, as a future projection, we consider that the experimental precision could reach the theoretical one in upcoming experiments. We do not include the muonic deuterium $2s$ hyperfine measurement due to the recently found tension between the experimental and theoretical determinations \cite{Kalinowski:2018vjm}. 
 
\begin{itemize}
 \item \textbf{Hydrogen H} The $1s$ HFS in hydrogen is one of the most accurate measurements ever made (see e.g.~\cite{Karshenboim:2005iy}). However, the precision of its theoretical description is limited by the presence of nuclear effects. Therefore, the best probe for new physics is actually the measurement of the $2s$ HFS, where we use the very precise $1s$ experimental measurement to quantify the finite size effects in the theoretical prediction. For this we use the theoretical analysis in \cite{PhysRevA.73.062503}, where analytical computation of the shift $8E(2s)-E(1s)$ (free of finite size effects) is used. 
 The precision here is limited by the latest experimental measurement \cite{PhysRevLett.102.213002}.
 \item  \textbf{Muonic Hydrogen $\mu$H}   For the muonic hydrogen $2s$ HFS we consider the experimental measurement of \cite{Pohl:2010zza,Antognini:1900ns} and compare it to the theoretical prediction presented in \cite{Peset:2016wjq}, where the main nuclear size effect, the two photon exchange contribution, has been computed from the very precise $1s$ measurement in regular hydrogen. 
  Finally, we consider the future proposed experiments that will measure the ground-state HFS in muonic hydrogen \cite{Adamczak:2012zz,Adamczak:2016pdb,Schmidt:2018kjc} with an uncertainty that could reach below the $\mu$eV. As in the case of regular hydrogen, since the bulk of the HFS theoretical uncertainty still comes from two-photon exchange effects, this new very precise measurement could be used to further reduce the uncertainty of the $2s$ theoretical prediction.  By getting rid of the experimental error in the two-photon-exchange \cite{Peset:2016wjq} we consider a feasible future improvement of the uncertainty of a factor $\sim$1/10.
 \end{itemize}


\begin{table}[!htbp]
\centering
\begin{tabular}{ccc|c|c}
\toprule
System  & \multicolumn{2}{c}{$2s$ Hyperfine }& \multicolumn{2}{c}{ }\\
\midrule
{}    & Exp. (MHz)   & Theo. (MHz)&$\sigma_{max} $  (Hz)& $\sigma_\text{proj} $ (Hz)\\
H   &177 .5568343(67) \cite{PhysRevLett.102.213002} &177. 5568382(3) & 6.7 & 0.3 \\
\bottomrule 
{}   & Exp. (meV)    & Theo. (meV)&$\sigma_{max} $  ($\mu$eV)& $\sigma_\text{proj} $ ($\mu$eV)\\
$\mu$H    & 22.8089(51) \cite{Pohl:2010zza,Antognini:1900ns}   &22.812(3)  & 5.1  &0.5  \\
\bottomrule 
System  & \multicolumn{2}{c}{$1s$ Hyperfine }& \multicolumn{2}{c}{ }\\
\midrule
$\mu$H    & --   & 182.62(2) & -- &20  \\
\bottomrule 
\end{tabular}
\caption{Table for the theoretical and experimental predictions of the hyperfine splitting in different muonic systems as well as in hydrogen. See the text for references to the theoretical computations and the origin of the projected uncertainty.}\label{table:HF}
\end{table}

\subsection{ Spin-independent dark forces}
In this section we discuss the sensitivity of  the measurements previously discussed to spin-independent dark forces, that is, forces mediated by either a new scalar or a vector.
We will present a comparison between this and other model independent laboratory probes. It is important to notice though that both spin independent and spin dependent new forces coupled to muons and protons are strongly constrained by astrophysical bounds such as the supernova SN1987A explosion \cite{astroMUON}.
\par
\begin{figure}[ht]
\begin{center}
\includegraphics[width=6.8 cm,height=5cm]{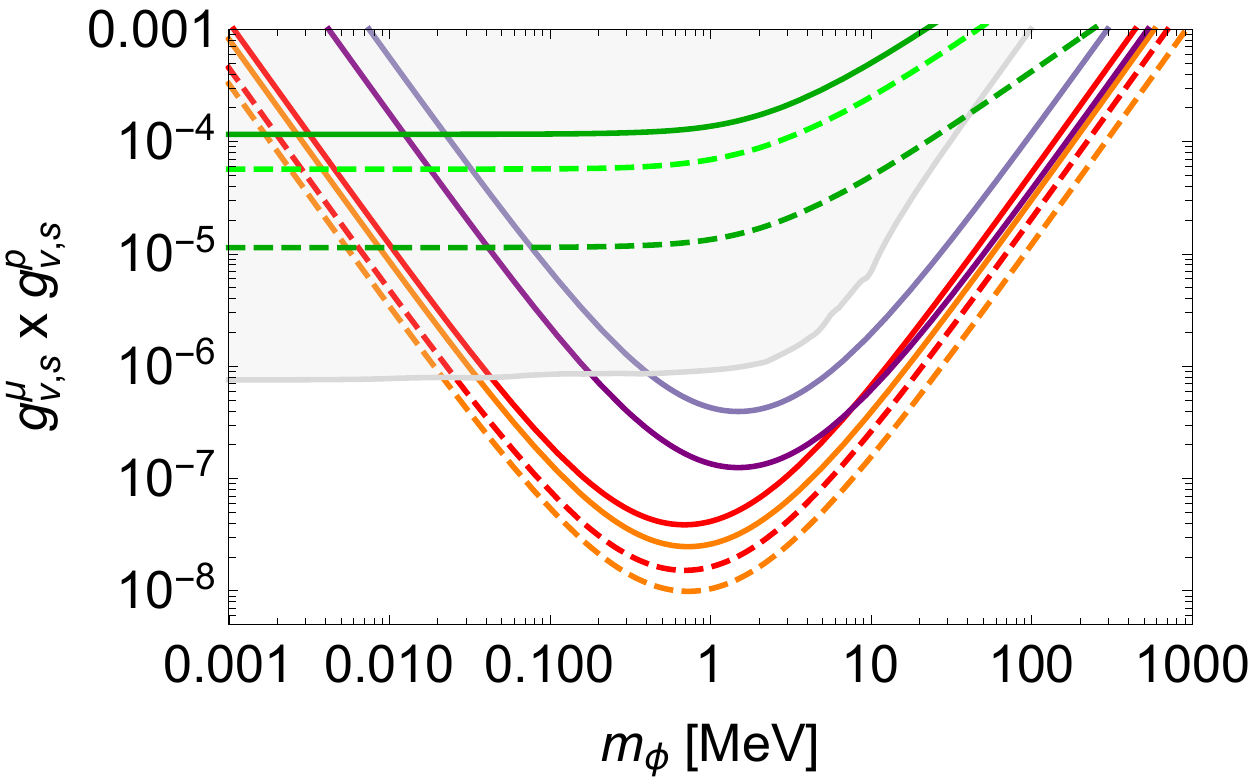}
\includegraphics[width=6.8 cm,height=5 cm]{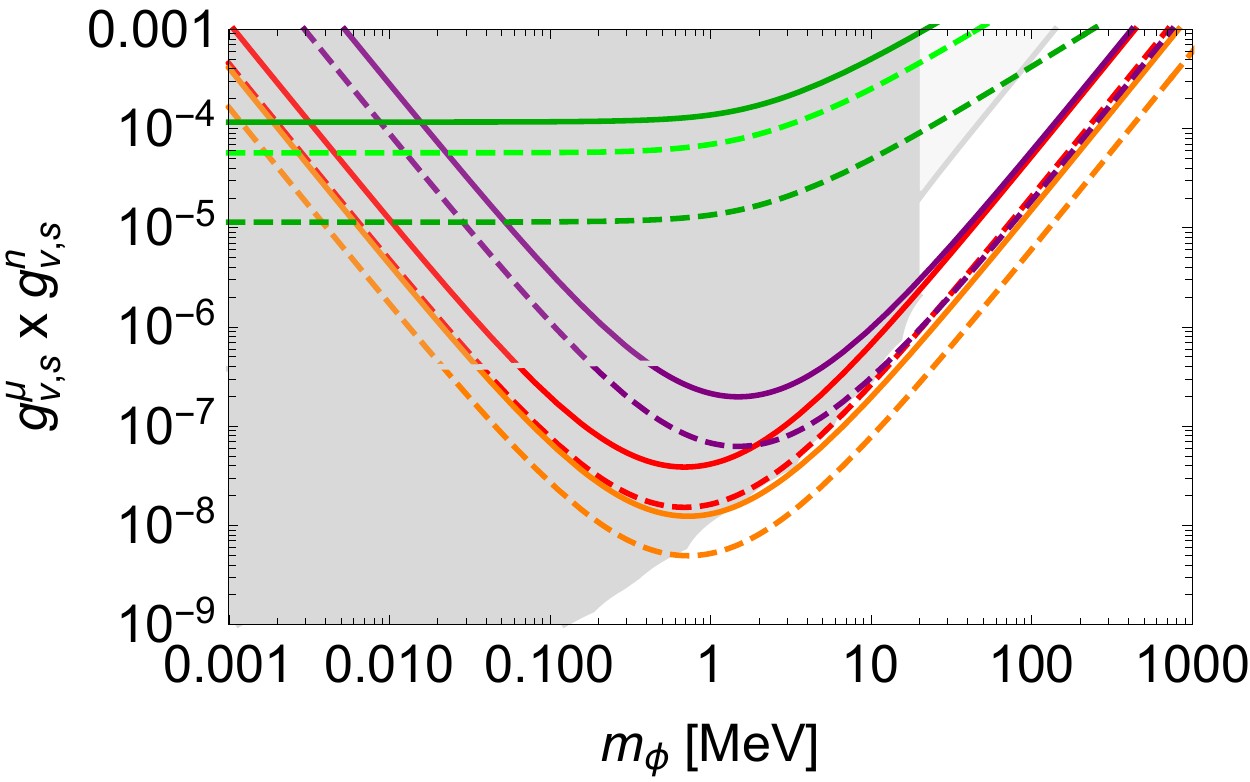}
\caption{\textit{Left plot}: constraints on the dimensionless coupling $g^{\mu} \times g^p$  as a
function of the scalar/vector mass $m_{\phi}$. All the lines correspond to the bounds coming from muonic atoms spectroscopy, see Table~\ref{table:LS} and Table~\ref{table:HF} for the experimental and theoretical precision. Dashed lines represent the future projections obtained in Tables~\ref{table:LS} and~\ref{table:HF}. The red/orange/purple curves represent the bound coming from $\mu H$/$\mu$D/$\mu^4$He Lamb shift measurements respectively. The green curves correspond to the bound coming from HFS measurements and apply only in the case of a vector mediator.
The two darker green are the constraints from the $2s$ HFS , while the lighter green is the projection for the $1s$ HFS considering the projection for the FAMU experiment, see Table~\ref{table:HF}. The light grey region corresponds to the region excluded by the  X-ray measurements on the  $3d_{5/2}-2p_{3/2}$ transitions in  muonic 24Mg
and 28Si \cite{vacchi}. \textit{Right plot}: the lines correspond to the same transitions as in the left panel with the difference  that we consider here a scenario where the new force couples both with proton and neutrons assuming $g^p=g^n=g^N$. Hence, the constraint is on  $g^{\mu} \times g^n$ where $g^n$ indicates the nucleon coupling. The dark grey region corresponds to   $g_{g2}^{\mu} \times g_{scattering}^n$, where $g_{g2}^{\mu}$ is the bound from the muon $g-2$ at 10$\sigma$ \cite{g2muon}. The neutron coupling $g^n$ is constrained by low-energy neutron scattering experiments up to masses $ \sim 20$~MeV, in the region of interest the strongest bound comes from \cite{leeb} (see \cite{atomic1} for an overview of these constraints).}\label{Fig:SI}
\end{center}
\end{figure}
\begin{figure}[ht]
\begin{center}
\includegraphics[width=6.8 cm,height=5cm]{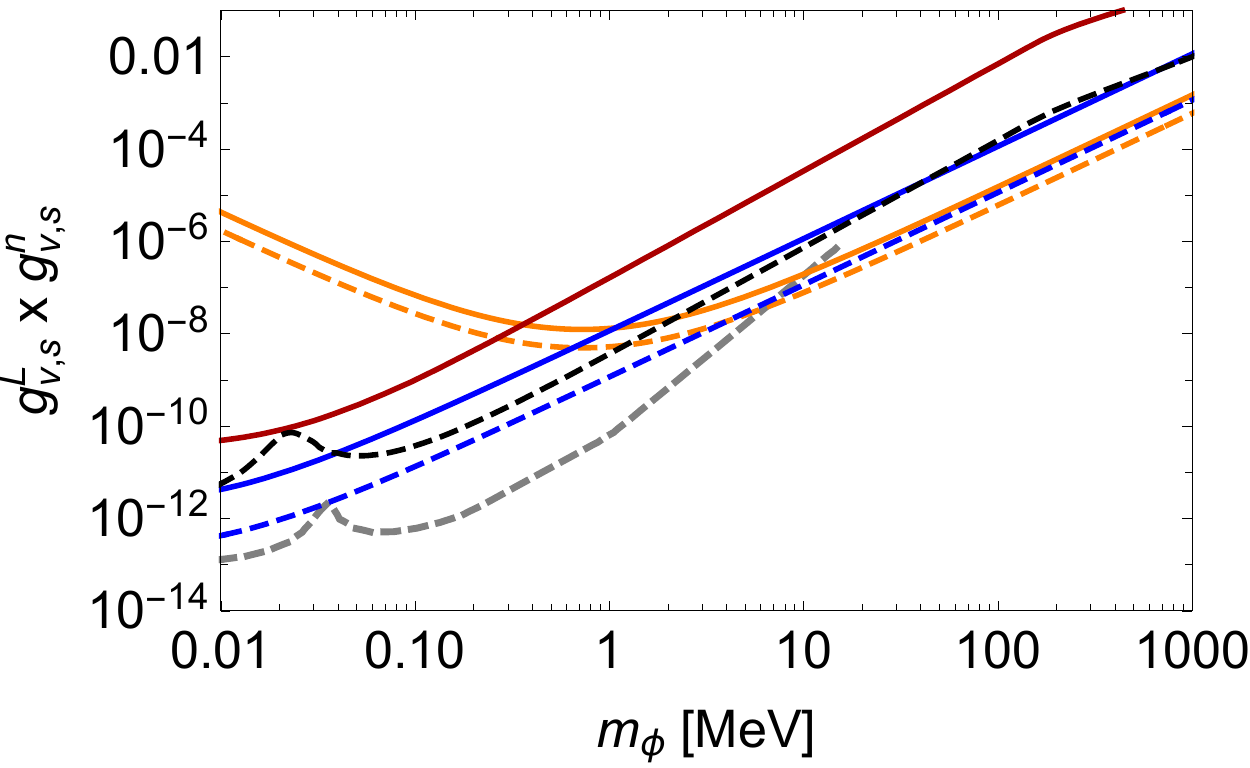}
\includegraphics[width=6.8 cm,height=5cm]{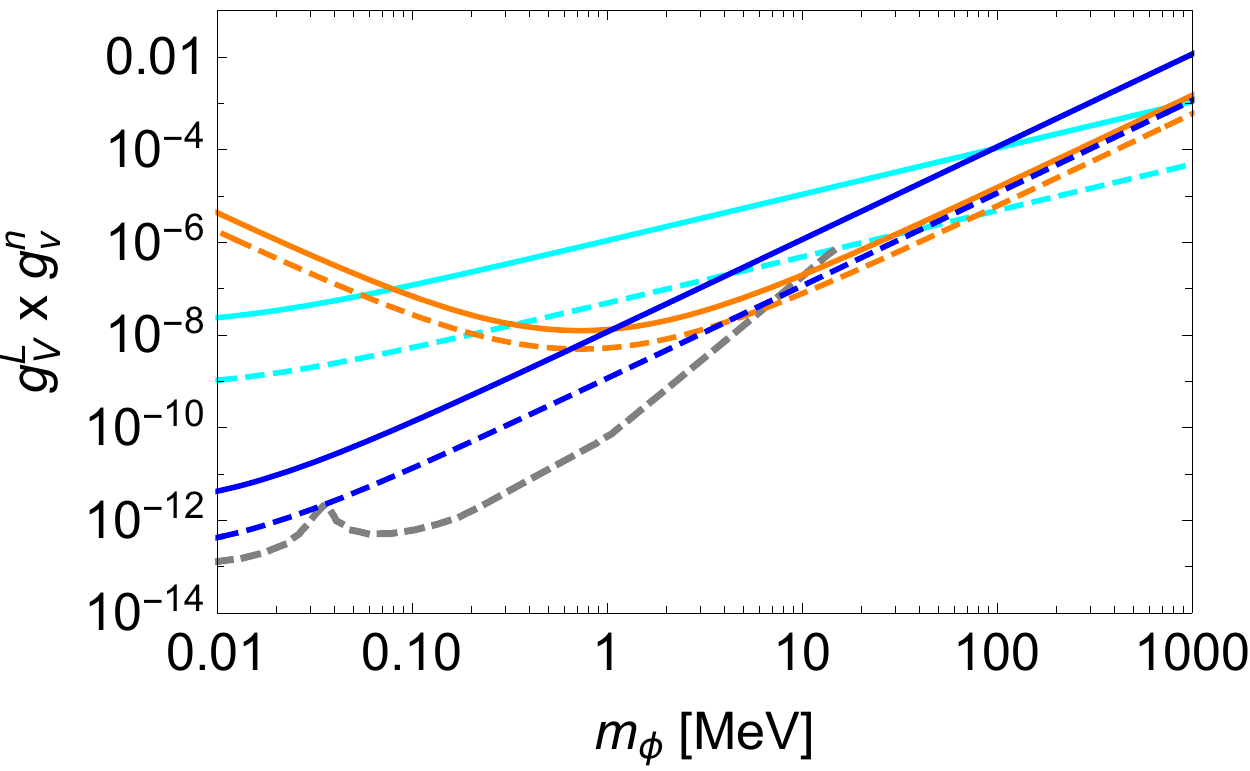}
\caption{\textit{Left plot}: comparison between the sensitivity reached by electronic and muonic atomic spectroscopy considering a new dark force mediated by either a scalar or a vector new particle. We assume $g^e=g^{\mu}$ and $g^p=g^n=g^N$. The orange line represents the bound from $\mu$D spectroscopy. The blue curve represents the updated bound from the latest H Lamb Shift measurement \cite{Bezginov:2019mdi}(considering also the muonic hydrogen measurement for the proton charge radius). Dashed lines correspond to future improvements. The grey dashed line represents the projection for the Yb IS measurement as proposed in \cite{kings1}, while the dashed black is the projection for the IS He$^4$ bound as explained in \cite{helium}. \textit{Right plot}: this plot applies only to the case where the new force is mediated by a vector new particle, since we also include (cyan lines) the bounds coming from $2s$ HFS in regular hydrogen (see Table~\ref{table:HF}).}\label{Fig:SIcomp}

\end{center}
\end{figure}

Our  main findings are captured in Fig.~\ref{Fig:SI} and Fig.~\ref{Fig:SIcomp}.
\begin{itemize}
\item Fig.~\ref{Fig:SI} presents the sensitivity of the Lamb shift measurements at PSI on $\mu$H, $\mu$D, and $\mu^4$He and for the vector case also of the HFS measurements.
 In the right panel  we assume no coupling to the neutrons, i.e. $g^n=0$, while on the left panel we consider a scenario with isospin preserving couplings  such that $g^n=g^p=g^N$.
In both cases, we notice that the $\mu$D is the measurement with the strongest sensitivity for masses above 1 MeV, both with respect to the other muonic measurements and to other indirect laboratory probes (grey regions), i.e. probes which do not depend on the production and subsequent decay of the new force mediator. 
The strongest  existing bound  on $ g^{\mu} \times g^p$ comes  from the measurements on  the  $3d_{5/2}-2p_{3/2}$ transitions in  muonic 24Mg
and 28Si \cite{vacchi} (grey region in the left panel of Fig.~\ref{Fig:SI}).
If the new particle couples also to neutrons, as it happens naturally in many new physics scenarios i.e. \cite{relaxion,atomic1}, the neutron coupling $g^n$ is  constrained by low-energy neutron scattering experiments up to masses $ \sim 20$~MeV. In the region of interest, the strongest bound comes from \cite{leeb} (see \cite{atomic1} for an overview of these constraints).
Hence, in the right panel of Fig.~\ref{Fig:SI} we present the bound from $g_{g2}^{\mu} \times g_{scattering}^n$ as a darker grey region ($g^{\mu}_{g2}$ is the 10$\sigma$ constraint on the muon coupling from the $g-2$ measurement \cite{g2muon} and $ g_{scattering}^n$ \cite{leeb}).
Even though the $\mu$D bound is only slightly stronger than the existing one, it is more solid than the neutron scattering due to several uncertainties in the latter one (see \cite{atomic1} for a brief discussion).
Finally, we notice that the bounds from HFS measurements (three green lines), valid only in the case of a vector mediator, are far less strong than the ones obtained by these old measurements. This is also true considering the projection for future improvements. This is easily explained by the suppression of the HFS contribution from a vectorial new force compared to the Lamb shift.
\item
In Fig.~\ref{Fig:SIcomp} we compare the bounds from muonic atoms to the bounds coming from other systems. The left panel presents the comparison of the reach of the $\mu$D measurement to the one of standard atoms considering both the hydrogen LS measurement (Table~\ref{table:LS}) and isotope shift (IS) measurements\footnote{In this case two approaches are viable, either the one of bounding new physics via $\sigma_{max}$ \cite{helium} also followed in this work or via the Kings plot measurements \cite{kings}.}.
Since all these measurements are sensitive to different couplings (electrons, protons, neutrons or protons) we assume  $ g^p=g^n=g^N$ and $g^e=g^{\mu}$.
 In Fig.~\ref{Fig:SIcomp} we present for the first time an updated bound from the LS measurement on standard hydrogen considering the latest measurement \cite{Bezginov:2019mdi} and the updated measurement of the proton charge radius from muonic hydrogen.
We find that this is, at the moment, the strongest bound from atomic physics spectroscopy in the light mass regime.
In the near future this could be overcome by the IS measurements on helium \cite{helium}.
An even stronger sensitivity could be reached by reaching  Yb measurements \cite{kings,kings1}, see  \cite{kings1} for a discussion on the treatment of non linearity. 
We see that the $\mu$D bound is the strongest among the existing ones for $m_{\phi} >1 $~MeV overcoming for $m_{\phi}>10$~MeV also the projections \cite{helium,kings1}. The right panel presents the comparison between atomic probes including also the $2s$ HFS sensitivity for regular hydrogen. This bound is valid only for the vector case.
We see that a future improvement of the experimental accuracy to match the theoretical one, could provide the best sensitivity among atomic probes for masses above 100~MeV.

\end{itemize}

\clearpage
\subsection{ Spin-dependent  dark forces}
\begin{figure}[ht]
\begin{center}
\includegraphics[width=6.8 cm,height=5cm]{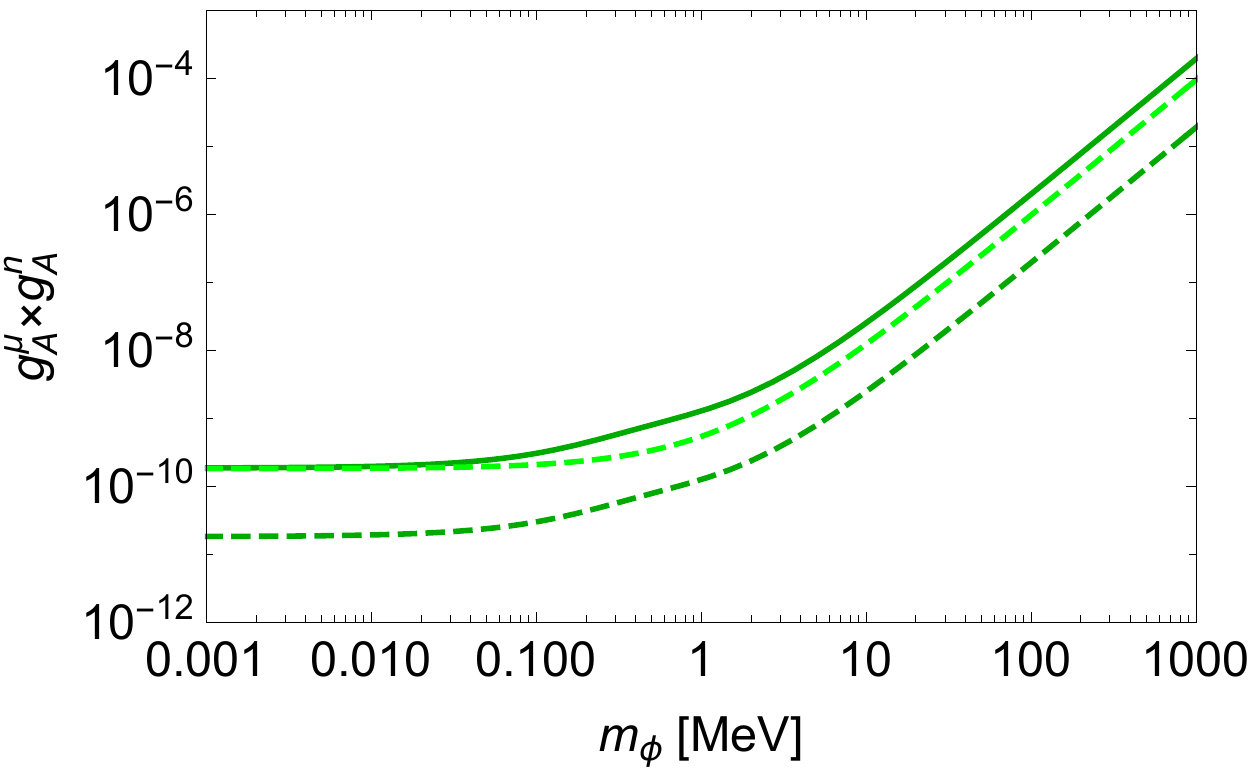}
\includegraphics[width=6.8 cm,height=5cm]{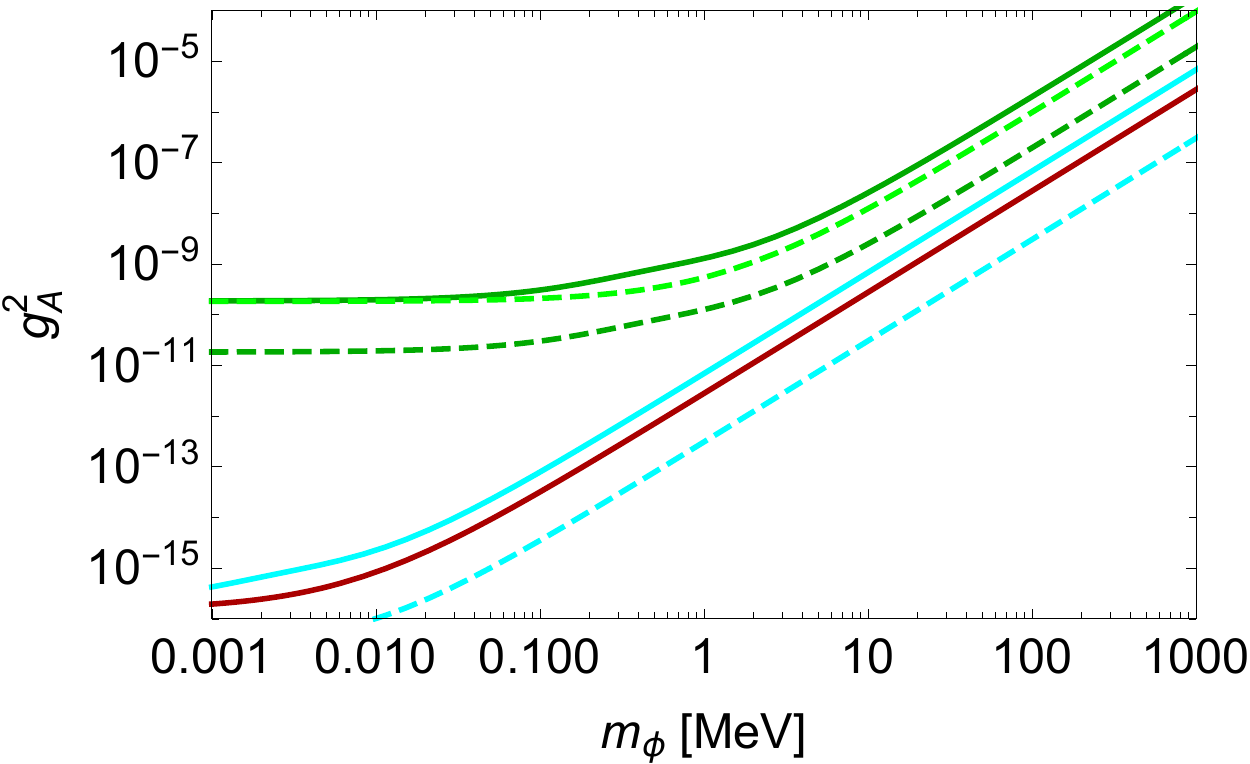}
\caption {\textit{Left plot:} Sensitivity to a new dark axial force coupled to protons and muons of the HFS measurement on $\mu$H. Dark green lines correspond to the $2s$ HFS and light green dashed the projection for the $1s$ HFS measurement proposed by the FAMU experiment, see Table~\ref{table:HF} for details and references. Dashed lines correspond to projected future improvements. \textit{Right plot:} Comparison between the muonic measurements and other HFS measurements on hydrogen and muonium $\mu e$ assuming that $g_{\mu}^A=g_p^A=g_e^A$.
}\label{Fig:SD}
\end{center}
\end{figure}

We first study the sensitivity to a new axial dark force. Such a scenario has been mainly studied in the context of  experimental anomalies such as the Beryllium anomaly  
(see for instance \cite{axial1}).
Fig.~\ref{Fig:SD} presents the sensitivity of HFS in muonic spectroscopy considering the present bound from the CREMA $2s$ HFS measurement in muonic hydrogen and the projections also for the future FAMU measurement of the $1s$ HFS.
In the right panel, the sensitivity of muonic spectroscopy is compared to the one of muonium HFS \cite{Karshenboim:2010cj} and, for the first time, to the hydrogen $2s$ HFS, assuming equal couplings to all fermions. We notice that the future hydrogen $2s$ HFS measurement will provide the strongest probe of such dark force among precision atomic spectroscopy. We checked that the $2s$ HFS measurement provides a stronger bound than the $1s$ HFS presented in \cite{Karshenboim:2010cg}: indeed, the sensitivity of the latter is limited by the theoretical uncertainty.

\begin{figure}[ht]
\begin{center}
\includegraphics[width=6.8 cm,height=5 cm]{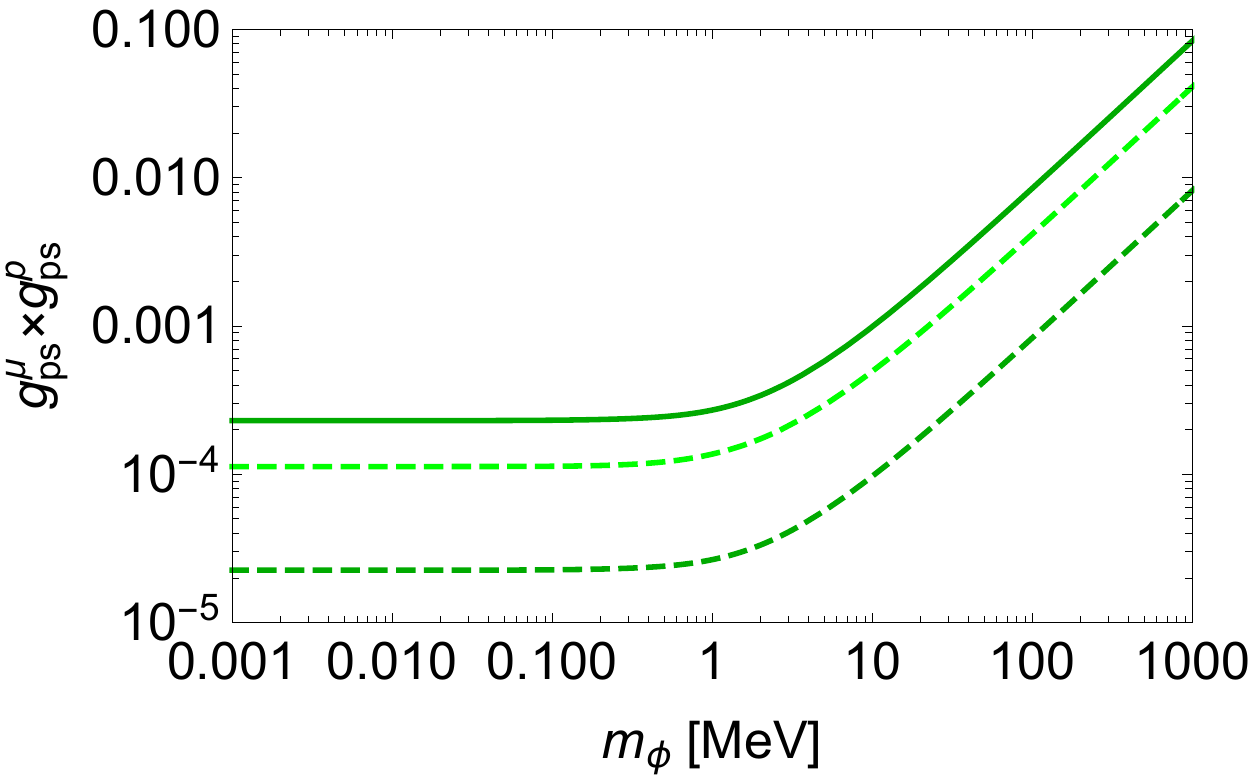}
\includegraphics[width=6.8 cm,height=5 cm]{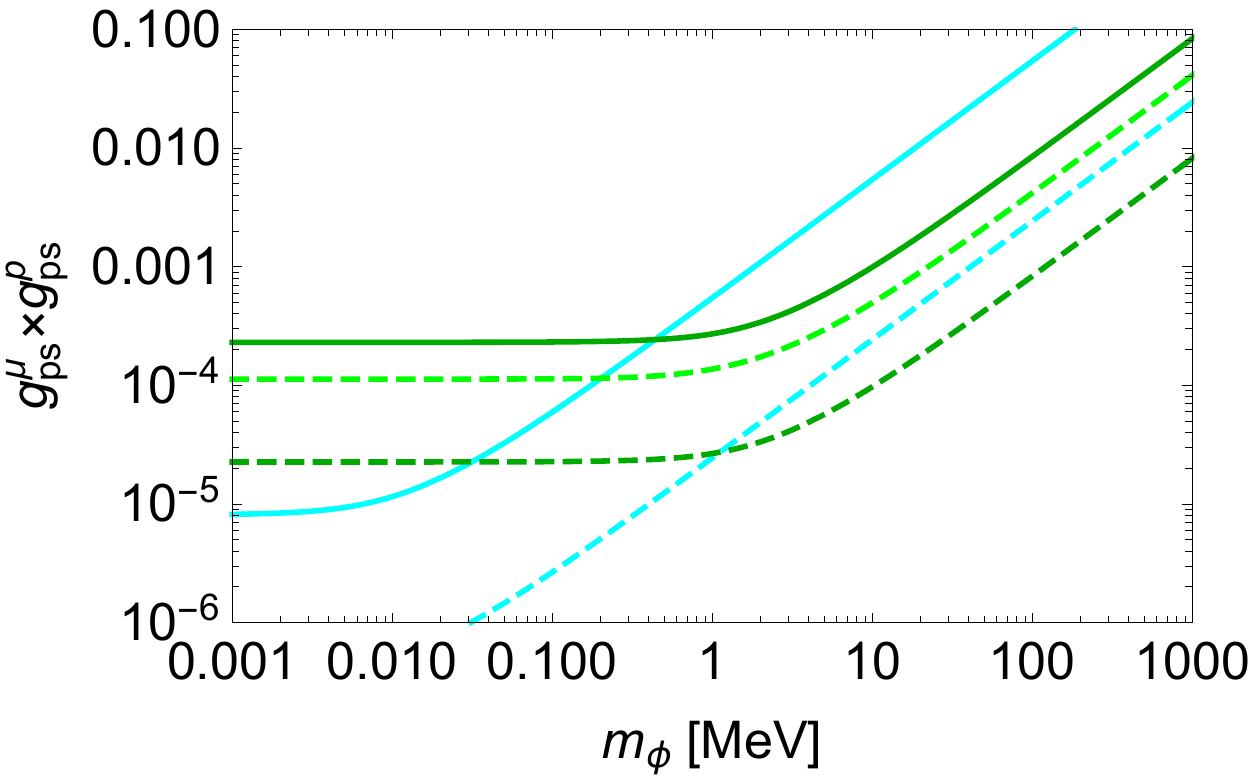}
\caption{ \textit{Left Plot}: Sensitivity to a light axion-like particle coupled to muons and protons of the $\mu$H HFS measurements and projections.
The dark green curves correspond to the $2s$ HFS present (solid) and future (dashed) bounds, while the light dashed green to the future $1s$ HFS FAMU measurement, see Table~\ref{table:HF}. \textit{Right plot}: we consider a scenario where $ g^e=g^{\mu} \frac{m_e}{m_{\mu}}$ and compare $\mu$ HFS measurements to the $2s$ HFS on regular hydrogen.}\label{Fig:SDcomp}
\end{center}
\end{figure}

Finally, we consider the case of a dark force mediated by a light pseudo-scalar, which are ubiquitous in many BSM scenarios (for a review \cite{axion1,axion2}). 
In the left panel of Fig.~\ref{Fig:SDcomp}, we can appreciate how the overall sensitivity of spectroscopy is very limited compared to other force carriers, the reason being the mass suppression of its leading potential.
 The right panel of Fig.~\ref{Fig:SDcomp} presents the comparison with the bound from the $2s$ HFS in regular hydrogen, considering a scenario where $ g^e=g^{\mu} \frac{m_e}{m_{\mu}}$. In this case, the muonic spectroscopy provides the strongest bound (also taking into account future projections) for sufficiently heavy particles ($m_\phi\gtrsim 1$~MeV).

\section{Conclusions}
Precision atomic spectroscopy offers a powerful and robust probe of the existence of new light dark sectors.
In this paper we studied the sensitivity of the Lamb Shift and Hyperfine splitting measurements in muonic atoms at PSI to  spin-independent and spin-dependent dark forces.
Among the atomic systems, muonic atoms are particularly sensitive to a new intermediate (MeV-GeV) scale.
This is a rather unique feature in the MeV-GeV region (other examples include the leptonic giromagnetic factor \cite{Odom:2006zz,g2e2008} or electric dipole moments \cite{Baron:2013eja}) since most of the other probes both at the precision (such as rare meson decays) and intensity frontier depend on the decay modes of the new particle. 
\par
Our main findings are the following:
\begin{itemize}
\item We develop the EFT that describes the leading effect of a new (pseudo-)vector or (pseudo-)scalar boson of \emph{any mass} at atomic energies. While this was already studied for light masses (of the order of the inverse size of the atom) the computation for intermediate and large masses of the one-loop matching coefficients is new. In our work this is particularly relevant for the description of the vector and pseudo-scalar bosons hyperfine splitting, where the leading effect is a one-loop effect rather than a tree level exchange. The EFT presented in Appendix~\ref{app:EFTforNP} however, is general and can be used to describe any effect at atomic energies.
\item  We find that the Lamb Shift  CREMA measurement on $\mu$D provides the strongest indirect  laboratory  bounds on the spin-independent bound $g^{\mu} \times g^n$ couplings for $ m_{\phi} > 1$ MeV region. 
Moreover, we find that for $m_{\phi} >1 $ MeV (where $\phi$ is the new particle mediating such a force) the present CREMA measurements supersede all existing atomic measurements including isotope shift measurements. Furthermore, for $m_{\phi} >10 $ MeV this is true also taking into account future projections \cite{helium,kings}.
 
\item We discussed for the first time the full expression and the importance of taking into account the contribution of a vector new force to the HFS: we show that the future HFS $2s$ measurement in Hydrogen would provide the strongest bound for masses $m_{\phi} >$ 100 MeV, see Fig.~\ref{Fig:SIcomp} .

\end{itemize}

\acknowledgments
 The work of CP is supported by the DFG within the Emmy
Noether Programme under grant DY130/1-1 and by the NSFC and the DFG
through the funds provided to the
Sino-German Collaborative Research Center TRR110 ``Symmetries and the
Emergence of Structure in QCD'' (NSFC
Grant No. 12070131001, DFG Project-ID 196253076 - TRR 110).

\appendix

\section{The EFT for new particle interactions}\label{app:EFTforNP}

We start by considering and additional (flavor preserving) interaction with the SM fermions that can be vector ($V$), scalar ($S$), pseudo-scalar ($P$) or axial-like ($A$) as in Eq.~\eqref{eq:NPLags}.

\subsection{The nonrelativistic EFT}\label{app:NREFT}
Since the fermions in a bound state move at very small relative velocity $v\ll 1$, these interactions can be matched onto a nonrelativistic effective field theory (NREFT). For QED this theory is NRQED \cite{Caswell:1985ui}. The precise shape of this theory will be different depending on the mass of the new particle $m_\phi$. We consider a light mass theory where $m_\phi\lesssim m_r\alpha$ and a heavy mass one where $m_\phi\sim m_r$. At order $g/m_r^2$, the fermion bilineal interactions are independent of the mediator mass
\begin{align}
\mathcal{L}_V^\text{NR}&=\psi^\dagger \left\{g_V V^0+\frac{i d_k g_V }{2m}(\boldsymbol{\nabla}{\bf V}+{\bf V}\boldsymbol{\nabla})-\frac{d_1g_V}{2m}\boldsymbol{\sigma}{\bf B}-\frac{d_2g_V}{8m^2}[\boldsymbol{\nabla},{\bf E}]-\frac{d_3 g_V}{8m^2}i\sigma^{ij}\{\nabla^i,E^j\}\right\}\psi\nn\\
&-(\psi\to \chi^c)\\
\mathcal{L}_A^\text{NR}&=-g_A\boldsymbol{\sigma }{\bf A}-\frac{g_A}{2m^2}d_8\{\boldsymbol{\nabla},\{\boldsymbol{\nabla},\boldsymbol{\sigma }\cdot{\bf A}\}\}+\frac{g_A}{4m^2}d_9 \{\boldsymbol{\sigma }\boldsymbol{\nabla},\boldsymbol{\nabla}{\bf A}\}-\frac{g_A}{4m^2}d_{10}[\boldsymbol{\nabla},(\boldsymbol{\sigma}\boldsymbol{\nabla}){\bf A}]\nn\\
&+\frac{g_A}{4m^2}d_{11}i (\boldsymbol{\nabla}\times {\bf A})\boldsymbol{\nabla}-\frac{g_A}{2m}d_{12}\{\boldsymbol{\sigma}\boldsymbol{\nabla},A^0\}-\frac{g_A}{8m^2}d_{12}[\boldsymbol{\sigma}\boldsymbol{\nabla},(\partial^0 A^0)]-(\psi\to \chi^c)\\
\mathcal{L}_S^\text{NR}&=\psi^\dagger \left\{g_S A^0+\frac{d_4 g_S}{8m^2}\{\boldsymbol{\nabla},\boldsymbol{\nabla}S+S\boldsymbol{\nabla}\}-\frac{id_5 g_S}{8m^2}\sigma^{ij}\{\nabla^i,(\nabla^jS)\}\right\}\psi+(\psi\to \chi^c)\\
\mathcal{L}_P^\text{NR}&=\psi^\dagger \left\{-\frac{d_6g_P}{2m}\boldsymbol{\sigma}[\boldsymbol{\nabla},P]+\frac{id_7 g_P}{8m^2}\boldsymbol{\sigma}\{\boldsymbol{\nabla},(\partial^0P)\}\right\}\psi+(\psi\to \chi^c)
\end{align}
where $\psi$ ($\chi^c$) is a two-component (anti-)quark field with mas $m$, ${\bf E}=-\boldsymbol{\nabla}V^0-\partial^0{\bf V}$, ${\bf B}=\boldsymbol{\nabla}\times {\bf V}$ and $\sigma^{ij}=-i/2[\sigma^i,\sigma^j]$. At leading order we find $d_i=1$. The vector-like Lagrangian reproduces the well known result for NRQED/QCD \cite{Caswell:1985ui,Pineda:2011dg,Beneke:2013jia}.

We also need to consider the four-fermion interactions
\begin{align}\label{def:LHM}
\mathcal{L}_X^\text{NR}&=-\frac{d_s^{(X)}}{m_1m_2}\psi^\dagger_1\psi_1\chi^{c\dagger}_2\chi^c_2+\frac{d_v^{(X)}}{m_1m_2}\psi^\dagger_1\boldsymbol{\sigma}\psi_1\chi^{c\dagger}_2\boldsymbol{\sigma}\chi_2^c,
\end{align}
with $X=V,A,S,P$. The Wilson coefficients in \eqref{def:LHM} do depend on the size of the mass of  the new particle. 

For a light mediator mass, the coefficients start at one loop for different fermion masses\footnote{For fermion-antifermion systems of equal mass, the annihilation diagrams will produce non-zero four-fermion Wilson coefficients already at tree level.} and will not produce the leading effect in any of the energy shifts we consider, thus they are beyond the scope of this work.

For heavy mediator masses $m_\phi\sim m_r$, the new particle is integrated out from the NREFT and its effects can only be seen through contact interactions as in \eqref{def:LHM}. We find that for the vector and scalar cases $d_s$ starts at tree level and $d_v$ at one loop while for the pseudo-scalar case both $d_v$ and $d_s$ start at one loop. We have computed the matching of these coefficients at one-loop with the help of the \texttt{FeynOnium} code \cite{Brambilla:2020fla}.
In the case of different mass components of the bound state, as was noted in \cite{Pineda:1998kj}, only the box-like diagrams contribute. We find:
\begin{align}
d_s^{(V)}&=g_V^{(1)}g_V^{(2)}\frac{m_1m_2}{2m_\phi^2}+\frac{g_V^{(1)}g_V^{(2)}Z\alpha m_2^2}{4(m_1^2-m_2^2)\pi}\left[1+2(1-2x^2)\ln(2x)\right.\nn\\
&\hspace*{5cm}\left.+\frac{2}{x^3}(1+2x^4)\sqrt{x^2-1}\ln(x+\sqrt{x^2-1})\right]+(m_1\leftrightarrow m_2),\\
d_v^{(V)}&=-\frac{g_V^{(1)}g_V^{(2)}Z\alpha m_1m_2}{4(m_1^2-m_2^2)\pi}\left[\ln\frac{m_1}{m_2}-\frac{4x^2}{3}\ln(2x)+\frac{4}{3x}(2+x^2)\sqrt{x^2-1}\ln(x+\sqrt{x^2}-1)\right]\nn\\
&\hspace*{.5cm}+(m_1\leftrightarrow m_2)\label{dvVm},
\end{align}
\begin{align}
d_s^{(A)}&=g_A^{(1)}g_A^{(2)}\frac{m_1m_2}{2m_\phi^2}+\frac{g_A^{(1)}g_A^{(2)}Z\alpha m_2^2}{6(m_1^2-m_2^2)\pi}\left[1-2\ln\frac{m_1}{m_2}-2\left(\frac{m_1^2}{m_2^2}+2x^2\right)\ln(2x)\right.\nn\\
&\left.\hspace*{5cm}-\frac{4}{x^3}(1-x^4)\sqrt{x^2-1}\ln(x+\sqrt{x^2}-1)\right]+(m_1\leftrightarrow m_2),\\
d_v^{(A)}&=-\frac{2g_A^{(1)}g_A^{(2)}Z\alpha m_1m_2}{(m_1^2-m_2^2)\pi}\left[-\frac{3}{4}\ln\frac{m_1}{m_2}-x^2\ln(2x)-\frac{1}{x}(1-x^2)\sqrt{x^2-1}\ln(x+\sqrt{x^2}-1)\right]\nn\\
&\hspace*{.5cm}+(m_1\leftrightarrow m_2),\label{dvAm}
\end{align}
\begin{align}
d_s^{(S)}&=-g_S^{(1)}g_S^{(2)}\frac{m_1m_2}{2m_\phi^2}-\frac{g_S^{(1)}g_S^{(2)}Z\alpha m_2^2}{(m_1^2-m_2^2)\pi}\left[\ln(2x)+\frac{1}{x^3}(1-x^2)\sqrt{x^2-1}\ln(x+\sqrt{x^2-1})\right]\nn\\
&\hspace*{.5cm}+(m_1\leftrightarrow m_2),\\
d_v^{(S)}&=\frac{g_S^{(1)}g_S^{(2)}Z\alpha m_1m_2}{2(m_1^2-m_2^2)\pi}\left[\ln\frac{m_1}{m_2}+\frac{4x^2}{3}\ln(2x)+\frac{4}{3x}(1-x^2)\sqrt{x^2-1}\ln(x+\sqrt{x^2-1})\right]\nn\\
&\hspace*{.5cm}+(m_1\leftrightarrow m_2),
\end{align}
\begin{align}
d_s^{(P)}&=0,\\
d_v^{(P)}&=\frac{g_P^{(1)}g_P^{(2)}Z\alpha m_1m_2}{3(m_1^2-m_2^2)\pi}\left[2x^2\ln(2x)-\frac{1}{x}(1+2x^2)\sqrt{x^2-1}\ln(x+\sqrt{x^2-1})\right]\nn\\
&\hspace*{.5cm}+(m_1\leftrightarrow m_2),\label{dvPm}
\end{align}
where $x\equiv m_\phi/(2m_1)$. We have checked that in the massless mediator limit, we recover the results obtained in \cite{Pineda:1998kj} for NRQED.
\subsection{The potential NREFT}\label{app:pNREFT}
Now, in order to obtain a sound EFT with well-defined power counting rules, we can obtain a potential EFT  by integrating out the soft scale $m_r\alpha$, where only ultrasoft modes are dynamical \cite{Pineda:1997bj,Brambilla:1999xf}. 

We compile here the full expressions for the position space potentials at order $g_X^2/m_r^2$ for a vector, axial-vector, scalar and pseudo-scalar mediator between a fermion of mass $m_1$ and an antifermion of mass $m_2$. It is known that there is a certain freedom on the choice of matching scheme from the NREFT to the potential-NREFT \cite{Peset:2015vvi}. We choose to do the matching off-shell, where potential loops cancel exactly and do not have to be considered.  
We find for the potentials $V_{XY}(r)$ with $X,Y=V,A,S,P$:
\begin{align}
V_{VV}(r)&=\frac{g_V^{(1)}g_V^{(2)}}{4\pi r}e^{-m_\phi r}+\frac{g_V^{(1)}g_V^{(2)}}{8\pi m_1m_2}\left[\frac{2 {\bf S^2}}{3}m_\phi^2\left(\frac{e^{-m_\phi r}}{r}-4\pi\delta^{(3)}(r)\right)+\frac{m_\phi}{2}  \left\{{\bf p}^2,e^{-m_\phi r}\right\}\right.\nn\\
&\left.+ \left\{{\bf p}^2,\frac{e^{-m_\phi r}}{2r}\right\}+\frac{m_\phi^3}{4}  e^{-m_\phi r}-\left(1-\frac{m_1 m_2}{4 m_r^2}\right)\frac{m_\phi^2 e^{-m_\phi r} }{r}+\left(1-\frac{m_1 m_2}{2 m_r^2}\right)2 \pi  \delta^{(3)}(r) \right.\nn\\
&\left.-\frac{\hat S_{12}(\hat{\bf r})}{2}  e^{-m_\phi r} \left(\frac{m_\phi^2}{3 r}+\frac{m_\phi}{r^2}+\frac{1}{r^3}\right)-e^{-m_\phi r} (m_\phi r+1)\frac{{\bf L}^2 }{r^3}\right.\nn\\
&\left.-\left(\frac{m_1 m_2}{2 m_r^2}+1\right) e^{-m_\phi r} (m_\phi r+1) \frac{{\bf LS}}{r^3}+\frac{m_1^2-m_2^2}{2 m_1 m_2} e^{-m_\phi r} (m_\phi r+1)\frac{{\bf LS^-} }{r^3 }\right]\nn\\
&+\frac{d_s^{(V)}+3d_v^{(V)}}{m_1m_2}\delta^{(3)}(r)-\frac{2d_v^{(V)}}{m_1m_2}{\bf S}^2\delta^{(3)}(r),
\end{align}
\begin{align}
V_{AA}(r)&=3\frac{g_A^{(1)}g_A^{(2)}}{4\pi r}e^{-m_\phi r}+\frac{g_A^{(1)}g_A^{(2)}}{2\pi r}e^{-m_\phi r}{\bf S}^2\nn\\
&-\frac{g_A^{(1)}g_A^{(2)}}{16\pi m_1m_2}\left\{(2 {\bf S}^2-3) \left[\frac{2 \left(m_1^2+m_2^2\right) }{m_1 m_2}\left\{{\bf p}^2,\frac{e^{-m_\phi r}}{2 r}\right\}\right.\right.\nn\\
&\left.\left.-m_\phi \text{acon}\left({\bf p}^2,e^{-m_\phi r}\right)+2e^{-m_\phi r} (m_\phi r+1)\frac{ {\bf L}^2 }{r^3}-\left(\frac{2}{3}-\frac{m_1 m_2}{2 m_r^2}\right)\frac{m_\phi^2 e^{-m_\phi r} }{r}\right.\right.\nn\\
&\left.\left. -\frac{1}{2} m_\phi^3 e^{-m_\phi r}+\left(\frac{5}{3}-\frac{m_1 m_2}{2 m_r^2}\right)4 \pi  \delta^{(3)}(r)\right]+\frac{m_1^2-m_2^2}{m_1m_2} e^{-m_\phi r} (m_\phi r+1)\frac{{\bf LS} }{r^3 }\right.\nn\\
&\left.+\frac{m_1^2+m_2^2}{m_1m_2} e^{-m_\phi r} (m_\phi r+1)\frac{{\bf LS^-} }{r^3}-\frac{8m_1 m_2}{m_r^2}\left({\bf S}_1\cdot {\bf p}\frac{e^{-m_\phi r}}{r}{\bf S}_2\cdot {\bf p}\right) \right.\nn\\
&\left.+\hat S_{12}(\hat{\bf r}) e^{-m_\phi r} \left(\frac{m_\phi^2}{3 r}+\frac{m_\phi}{r^2}+\frac{1}{r^3}\right)\right]+\frac{d_s^{(A)}+3d_v^{(A)}}{m_1m_2}\delta^{(3)}(r)-\frac{2d_v^{(A)}}{m_1m_2}{\bf S}^2\delta^{(3)}(r),
\end{align}
\begin{align}
V_{SS}(r)&=-\frac{g_S^{(1)}g_S^{(2)}}{4\pi r}e^{-m_\phi r}-\frac{g_S^{(1)}g_S^{(2)}}{8\pi m_1m_2}\left[- \left(1-\frac{m_1 m_2}{2 m_r^2}\right) \left\{{\bf p}^2,\frac{e^{-m_\phi r}}{r}\right\}-\frac{ m_\phi}{2} \left\{{\bf p}^2,e^{-m_\phi r}\right\}\right.\nn\\
&\left.+e^{-m_\phi r} (m_\phi r+1)\frac{{\bf L}^2 }{r^3}+\left(1-\frac{m_1 m_2}{2 m_r^2}\right)e^{-m_\phi r} (m_\phi r+1) \frac{{\bf LS} }{r^3}\right.\nn\\
&\left.+\frac{m_1^2-m_2^2}{m_1m_2} e^{-m_\phi r} (m_\phi r+1)\frac{{\bf LS^-} }{r^3}-\left(1-\frac{m_1 m_2}{2 m_r^2}\right)\frac{m_\phi^2 e^{-m_\phi r} }{2 r}\right.\nn\\
&\left.+4 \pi  \delta^{(3)}(r) \left(1-\frac{m_1 m_2}{4 m_r^2}\right)-\frac{1}{4} m_\phi^3 e^{-m_\phi r}\right]+\frac{d_s^{(S)}+3d_v^{(S)}}{m_1m_2}\delta^{(3)}(r)-\frac{2d_v^{(S)}}{m_1m_2}{\bf S}^2\delta^{(3)}(r),
\end{align}
\begin{align}
V_{PP}(r)&=\frac{g_P^{(1)}g_P^{(2)}}{16\pi m_1m_2}\left[\frac{m_\phi^2 e^{-m_\phi r}}{r}-4 \pi  \delta^{(3)}(r)\right](1-\frac{2}{3}{\bf S}^2)\nn\\
&-\frac{g_P^{(1)}g_P^{(2)}}{16\pi m_1m_2}\hat S_{12}(\hat{\bf r}) e^{-m_\phi r} \left(\frac{m_\phi^2}{3 r}+\frac{m_\phi}{r^2}+\frac{1}{r^3}\right)+\frac{d_s^{(P)}+3d_v^{(P)}}{m_1m_2}\delta^{(3)}(r)\nn\\
&-\frac{2d_v^{(P)}}{m_1m_2}{\bf S}^2\delta^{(3)}(r),
\end{align}
where $\hat S_{ij}(\hat{\bf r})=-4({\bf S}_i\cdot {\bf S}_j)+12 ({\bf S}_i\cdot \hat{\bf r})({\bf S}_j\cdot \hat{\bf r})$. We remark that these are the full potentials up to order $g_X^2/m^2$. They complete the results obtained in \cite{Fadeev:2018rfl} where all the potentials proportional to $1/m_1^2$ and $1/m_2^2$ and ${\bf L}^2$ (or the equivalent potential loops in the on-shell matching scheme) are missing. The vector potential in the massless mediator limit $m_\phi\to 0$ reproduces the well-known pNRQED result \cite{Pineda:1997bj,Peset:2015zga}. 

For a heavy mass mediator the potentials can be fully expressed in terms of the Wilson coefficients in Eqs.~\eqref{dvVm}-\eqref{dvPm} and read
\begin{align}
V_{XX}(r)=\frac{d_s^{(X)}+3d_v^{(X)}}{m_1m_2}\delta^{(3)}(r)-\frac{2d_v^{(X)}}{m_1m_2}{\bf S}^2\delta^{(3)}(r).
\end{align}
This completes the pNREFT that describes the effect of a new dark force at atomic energies.
\section{Expectation values}\label{app:EVs}

In order to compute energy shifts from the potentials in appendix~\ref{app:pNREFT} we need to compute the expectation values of the different operators. We give here some general expressions for their computation.

Spin-like expectation values \cite{Kiyo:2014uca}:
\begin{align}
&\langle {\bf S}^2\rangle=s(s+1),\quad \quad \langle {\bf LS}\rangle=\frac{1}{2}(j(j+1)-l(l+1)-s(s+1)),\nn\\
&\langle {\bf LS_i}\rangle=\frac{1}{2}(j(j+1)-l(l+1)-s_i(s_i+1)),\quad\quad\langle {\bf LS^-}\rangle=2\langle {\bf LS_1}\rangle-\langle {\bf LS}\rangle,\nn\\
&\langle\hat S_{ij}(\hat{\bf r})\rangle=4\frac{2l(l+1)s(s+1)-3\langle {\bf LS}\rangle-6\langle {\bf LS}\rangle^2}{(2l-1)(2l+3)}.
\end{align}

We also compile here a general expression for the relevant radial expectation values:
\begin{align}
&\langle \frac{e^{-m_\phi r}}{r^x}\rangle=\frac{4 (n-l-1)! \left(\frac{a_0 n}{2}\right)^{3-x} }{\left(a_0^3 n^4\right) (l+n)!}\sum _{k=0}^{\infty } \frac{\left(-\frac{ m_\phi a_0 n}{2} \right)^k \Gamma (k+2 l-x+3) \Gamma (-k-l+n+x-2)^2}{\Gamma (k+1) \Gamma (-k+x-1)^2 \Gamma (n-l)^2}\nn\\
&\hspace*{1.25cm}\times \, _3F_2(l-n+1,l-n+1,k+2 l-x+3;k+l-n-x+3,k+l-n-x+3;1),\nn\\
&\langle {\bf p}^2,\frac{e^{-m_\phi r}}{r^x}\rangle=\frac{4}{a_0}\langle \frac{e^{-m_\phi r}}{r^{x+1}}\rangle-\frac{2}{n^2a_0^2}\langle \frac{e^{-m_\phi r}}{r^x}\rangle.
\end{align}

\bibliographystyle{JHEP}
\bibliography{lightatoms-bib}

\end{document}